\documentclass[sigconf]{acmart}
\AtBeginDocument{%
\providecommand\BibTeX{{%
\normalfont B\kern-0.5em{\scshape i\kern-0.25em b}\kern-0.8em\TeX}}}

\newif\ifnotes
\notesfalse 

\definecolor{purplecolor}{RGB}{128, 0, 128}
\definecolor{grey}{RGB}{128, 128, 128}

\copyrightyear{2021} 
\acmYear{2021} 
\setcopyright{rightsretained} 
\acmConference[CHI '21]{CHI Conference on Human Factors in Computing Systems}{May 8--13, 2021}{Yokohama, Japan}
\acmBooktitle{CHI Conference on Human Factors in Computing Systems (CHI '21), May 8--13, 2021, Yokohama, Japan}
\acmDOI{10.1145/3411764.3445211}
\acmISBN{978-1-4503-8096-6/21/05}

\begin{document}
\title[Viral Visualizations]{Viral Visualizations: How Coronavirus Skeptics Use Orthodox Data Practices to Promote Unorthodox Science Online}


\author{Crystal Lee}
\email{crystall@mit.edu}
\orcid{0000-0001-6672-9118}
\affiliation{%
  \institution{Massachusetts Institute of Technology}
  \city{Cambridge}
  \state{MA}
  \country{USA}
}

\author{Tanya Yang}
\email{tanyang@mit.edu}
\affiliation{%
  \institution{Massachusetts Institute of Technology}
  \city{Cambridge}
  \state{MA}
  \country{USA}
}

\author{Gabrielle Inchoco}
\email{ginchoco@wellesley.edu}
\affiliation{%
  \institution{Wellesley College}
  \city{Wellesley}
  \state{MA}
  \country{USA}
}

\author{Graham M. Jones}
\email{gmj@mit.edu}
\affiliation{%
  \institution{Massachusetts Institute of Technology}
  \city{Cambridge}
  \state{MA}
  \country{USA}
}

\author{Arvind Satyanarayan}
\email{arvindsatya@mit.edu}
\affiliation{%
  \institution{Massachusetts Institute of Technology}
  \city{Cambridge}
  \state{MA}
  \country{USA}
}

\renewcommand{\shortauthors}{Lee, Yang, Inchoco, Jones, and Satyanarayan}


\begin{abstract}
Controversial understandings of the coronavirus pandemic have turned data visualizations into a battleground. Defying public health officials, coronavirus skeptics on US social media spent much of 2020 creating data visualizations showing that the government’s pandemic response was excessive and that the crisis was over. This paper investigates how pandemic visualizations circulated on social media, and shows that people who mistrust the scientific establishment often deploy the same rhetorics of data-driven decision-making used by experts, but to advocate for radical policy changes. Using a quantitative analysis of how visualizations spread on Twitter and an ethnographic approach to analyzing conversations about COVID data on Facebook, we document an epistemological gap that leads pro- and anti-mask groups to draw drastically different inferences from similar data. Ultimately, we argue that the deployment of COVID data visualizations reflect a deeper sociopolitical rift regarding the place of science in public life.
\end{abstract}



\maketitle


\section{Introduction}

Throughout the coronavirus pandemic, researchers have held up the crisis as a ``breakthrough moment'' for data visualization research \cite{shneiderman_data_2020}: John Burn-Murdoch’s line chart comparing infection rates across countries helped millions of people make sense of the pandemic’s scale in the United States \cite{forrest_how_2020}, and even top Trump administration officials seemed to rely heavily on the Johns Hopkins University COVID data dashboard \cite{mazza_top_2020}. Almost every US state now hosts a data dashboard on their health department website to show how the pandemic is unfolding. However, despite a preponderance of evidence that masks are crucial to reducing viral transmission  \cite{yi-fong_su_masks_2020,cdc_coronavirus_2020,chu_physical_2020}, protestors across the United States have argued for local governments to overturn their mask mandates and begin reopening schools and businesses. A pandemic that affects a few, they reason, should not impinge on the liberties of a majority to go about life as usual. To support their arguments, these protestors and activists have created thousands of their own visualizations, often using the same datasets as health officials.

This paper investigates how these activist networks use rhetorics of scientific rigor to oppose these public health measures. Far from ignoring scientific evidence to argue for individual freedom, anti-maskers often engage deeply with public datasets and make what we call ``\textbf{counter-visualizations}''---visualizations using orthodox methods to make unorthodox arguments---to challenge mainstream narratives that the pandemic is urgent and ongoing. By asking community members to ``follow the data,'' these groups mobilize data visualizations to support significant local changes. 

We examine the circulation of COVID-related data visualizations through both quantitative and qualitative methods. First, we conduct a quantitative analysis of close to half a million tweets that use data visualizations to talk about the pandemic. We use network analysis to identify communities of users who retweet the same content or otherwise engage with one another (e.g., maskers and anti-maskers). We process over 41,000 images through a computer vision model trained by Poco and Heer \cite{poco_reverseengineering_2017}, and extract feature embeddings to identify clusters and patterns in visualization designs. The academic visualization research community has traditionally focused on mitigating chartjunk and creating more intuitive visualization tools for use by non-experts; \emph{better} visualizations, researchers argue, would aid public understanding of data-driven phenomena. However, we find that anti-mask groups on Twitter often create polished counter-visualizations that would not be out of place in scientific papers, health department reports, and publications like the \emph{Financial Times}. 

Second, we supplement this quantitative work with a six month-long observational study of anti-mask groups on Facebook. The period of this study, March to September 2020, was critical as it spanned the formation and consolidation of these groups at the pandemic’s start. Quantitative analysis gives us an overview of what online discourse about data and its visual representation looks like on Twitter both within and outside anti-mask communities. Qualitative analysis of anti-mask groups gives us an interactional view of how these groups leverage the language of scientific rigor---being critical about data sources, explicitly stating analytical limitations of specific models, and more---in order to support ending public health restrictions despite the consensus of the scientific establishment. Our data analysis evolved as these communities did, and our methods reflect how these users reacted in real time to the kaleidoscopic nature of pandemic life. As of this writing, Facebook has banned some of the groups we studied, who have since moved to more unregulated platforms (Parler and MeWe).

While previous literature in visualization and science communication has emphasized the need for data and media literacy as a way to combat misinformation \cite{guess_digital_2020, scheufele_science_2019,fontichiaro_why_2016}, this study finds that anti-mask groups practice a form of data literacy in spades. Within this constituency, unorthodox viewpoints do not result from a deficiency of data literacy; sophisticated practices of data literacy are a means of consolidating and promulgating views that fly in the face of scientific orthodoxy. Not only are these groups prolific in their creation of counter-visualizations, but they leverage data and their visual representations to advocate for and enact policy changes on the city, county, and state levels. 

As we shall see throughout this case study, anti-mask communities on social media have defined themselves \emph{in opposition} to the discursive and interpretive norms of the mainstream public sphere (e.g., against the ``lamestream media''). In media studies, the term ``counterpublic'' describes constituencies that organize themselves in opposition to mainstream civic discourse, often by agentively using communications media~\cite{downey_new_2003}. In approaching anti-maskers as a counterpublic (a group shaped by its hostile stance toward mainstream science), we focus particular attention on one form of agentive media production central to their movement: data visualization. We define this counterpublic’s visualization practices as ``counter-visualizations'' that use orthodox scientific methods to make unorthodox arguments, beyond the pale of the scientific establishment. Data visualizations are not a neutral window onto an observer-independent reality; during a pandemic, they are an arena of political struggle.

Among other initiatives, these groups argue for open access to government data (claiming that CDC and local health departments are not releasing enough data for citizens to make informed decisions), and they use the language of data-driven decision-making to show that social distancing mandates are both ill-advised and unnecessary. In these discussions, we find that anti-maskers think carefully about the grammar of graphics by decomposing visualizations into layered components (e.g., raw data, statistical transformations, mappings, marks, colors). Additionally, they debate how each component changes the narrative that the visualization tells, and they brainstorm alternate visualizations that would better enhance public understanding of the data. This paper empirically shows how COVID anti-mask groups use data visualizations to argue that the US government’s response (broadly construed) to the pandemic is overblown, and that the crisis has long been over. 

These findings suggest that the ability for the scientific community and public health departments to better convey the urgency of the US coronavirus pandemic may not be strengthened by introducing more downloadable datasets, by producing ``better visualizations'' (e.g., graphics that are more intuitive or efficient), or by educating people on how to better interpret them. This study shows that there is a fundamental epistemological conflict between maskers and anti-maskers, who use the same data but come to such different conclusions. As science and technology studies (STS) scholars have shown, data is not a neutral substrate that can be used for good or for ill \cite{bowker_sorting_2000, gitelman_raw_2013, porter_trust_1995}. Indeed, anti-maskers often reveal themselves to be more sophisticated in their understanding of how scientific knowledge is socially constructed than their ideological adversaries, who espouse naive realism about the ``objective'' truth of public health data. Quantitative data is culturally and historically situated; the manner in which it is collected, analyzed, and interpreted reflects a deeper narrative that is bolstered by the collective effervescence found within social media communities. Put differently, there is no such thing as dispassionate or objective data analysis. Instead, there are stories: stories shaped by cultural logics, animated by personal experience, and entrenched by collective action. This story is about how a public health crisis---refracted through seemingly objective numbers and data visualizations---is part of a broader battleground about scientific epistemology and democracy in modern American life. 


\section{Related Work}


\subsection{Data and visualization literacies}

There is a robust literature in computer science on data and visualization literacy, where the latter often refers to the ability of a person ``to comprehend and interpret graphs'' \cite{lee_vlat_2017} as well as the ability to create visualizations from scratch~\cite{borner_data_2019}. Research in this area often includes devising methods to assess this form of literacy \cite{lengler_identifying_2006,boy_principled_2014,alper_visualization_2017}, to investigate how people (mis)understand visualizations \cite{borner_investigating_2016,ellis_promoting_2018,ellis_studying_2018}, or to create systems that help a user improve their understanding of unfamiliar visualizations~\cite{stoiber_visualization_2019,ruchikachorn_learning_2015,tanahashi_study_2016,alberda_covid-19_2020,adar_communicative_2020,borner_data_2019}. Evan Peck et al. \cite{peck_data_2019} have responded to this literature by showing how a \textit{``complex tapestry of motivations, preferences, and beliefs} [impact] \textit{the way that participants} [prioritize] \textit{data visualizations,''} suggesting that researchers need to better understand users’ social and political context in order to design visualizations that speak powerfully to their personal experience. 

Linguistic anthropologists have shown that ``literacy'' is not just the ability to encode and decode written messages. The skills related to reading and writing are historically embedded, and take on different meanings in a given social context depending on who has access to them, and what people think they should and shouldn't be used for. As a consequence of such contingencies, these scholars view literacy as multiple rather than singular, attending to the impact of local circumstances on they way members of any given community practice literacy and construe its value~\cite{ahearn_literacy_2004}. Thus, media literacy, is not simply about understanding information, but about being able to actively leverage it in locally relevant social interactions~\cite{hobbs_exploring_2016}. Building on these traditions, we do not normatively assess anti-maskers' visualization practices against a prescriptivist model of what literacy \emph{should} be (according to, say, experts in human-computer interaction), but rather seek to describe what those practices actually look like in locally relevant contexts.

As David Buckingham \cite{buckingham_fake_2017} has noted, calls for increased literacy have often become a form of wrong-headed solutionism that posits education as the fix to all media-related problems. danah boyd \cite{boyd_you_2018} has documented, too, that calling for increased media literacy can often backfire: the instruction to ``question more'' can lead to a weaponization of critical thinking and increased distrust of media and government institutions. She argues that calls for media literacy can often frame problems like fake news as ones of personal responsibility rather than a crisis of collective action. Similarly, Francesca Tripodi \cite{tripodi_searching_nodate} has shown how evangelical voters do not vote for Trump because they have been ``fooled'' by fake news, but because they privilege the personal study of primary sources and have found logical inconsistencies not in Trump's words, but in mainstream media portrayals of the president. As such, Tripodi argues, media literacy is not a means of fighting ``alternative facts.'' Christopher Bail et al. \cite{bail_exposure_2018} have also shown how being exposed to more opposing views can actually increase political polarization. 

Finally, in his study of how climate skeptics interpret scientific studies, Frank Fischer \cite{fischer_knowledge_2019} argues that increasing fact-checking or levels of scientific literacy is insufficient for fighting alternative facts. ``While fact checking is a worthy activity,'' he says, ``we need to look deeper into this phenomenon to find out what it is about, what is behind it.'' Further qualitative studies that investigate how ideas are culturally and historically situated, as the discussion around COVID datasets and visualizations are manifestations of deeper political questions about the role of science in public life. 


\subsection{Critical approaches to visualization}

Historians, anthropologists, and geographers have long shown how visualizations---far from an objective representation of knowledge---are often, in fact, representations of power~\cite{harley_deconstructing_1989, teng_taiwans_2004,kitchin_introductory_2011,mueggler_paper_2011}. To address this in practice, feminist cartographers have developed quantitative GIS methods to describe and analyze differences across race, gender, class, and space, and these insights are then used to inform policymaking and political advocacy~\cite{hanson_geography_1992,mclafferty_counting_1995,kwan_feminist_2002}. Best practices in data visualization have often emphasized reflexivity as a way to counter the power dynamics and systemic inequalities that are often inscribed in data science and design~\cite{dignazio_data_2020,costanza-chock_design_2020,correll_ethical_2019}. Central to this practice is articulating what is excluded from the data \cite{onuoha_library_2016,buolamwini_gender_2018,noble_algorithms_2018}, understanding how data reflect situated knowledges rather than objective truth \cite{haraway_situated_1988,christin_daguerreotypes_2016,so_cartographers_2020}, and creating alternative methods of analyzing and presenting data based on anti-oppressive practices~\cite{kelly_mapping_2019,anti-eviction_mapping_project_dislocation_2019,data_for_black_lives_data_2020}. Researchers have also shown how interpreting data visualizations is a fundamentally social, narrative-driven endeavor~\cite{hullman_visualization_2011,peck_data_2019,hackett_social_2008}. By focusing on a user’s contextual experience and the communicative dimensions of a visualization, computer scientists have destabilized a more traditional focus on improving the technical components of data visualization towards understanding how users interpret and use them~\cite{viegas_communication-minded_2006,maltese_research_2015,lee_how_2016}.

Critical, reflexive studies of data visualization are undoubtedly crucial for dismantling computational systems that exacerbate existing social inequalities. Research on COVID visualizations is already underway: Emily Bowe et al. \cite{bowe_learning_2020} have shown how visualizations reflect the unfolding of the pandemic at different scales; Alexander Campolo \cite{campolo_flattening_2020} has documented how the pandemic has produced new forms of visual knowledge, and Yixuan Zhang et al. \cite{zhang_mapping_2020} have mapped the landscape of COVID-related crisis visualizations. This paper builds on these approaches to investigate the epistemological crisis that leads some people to conclude that mask-wearing is a crucial public health measure and for others to reject it completely. 

Like data feminists, anti-mask groups similarly identify problems of political power within datasets that are released (or otherwise withheld) by the US government. Indeed, they contend that the way COVID data is currently being collected is non-neutral, and they seek liberation from what they see as an increasingly authoritarian state that weaponizes science to exacerbate persistent and asymmetric power relations. This paper shows that more critical approaches to visualization are necessary, and that the frameworks used by these researchers (e.g., critical race theory, gender analysis, and social studies of science) are crucial to disentangling how anti-mask groups mobilize visualizations politically to achieve powerful and often horrifying ends. 


\section{Methods} 

This paper pairs a quantitative approach to analyzing Twitter data (computer vision and network analysis) with a qualitative approach to examining Facebook Groups (digital ethnography). Drawing from social media scholarship that uses mixed-methods approaches  to examine how users interact with one another \cite{burgess_mapping_2016, moats_quali-quantitative_2018, kiesling_interactional_2018, berriche_internet_2020, arif_acting_2018}, this paper engages with work critical of quantitative social media methods \cite{baym_data_2013, wu_platform_2020} by demonstrating how interpretive analyses of social media discussions and computational techniques can be mutually re-enforcing. In particular, we leverage quantitative studies of social media that use network analysis to understand political polarization \cite{arif_acting_2018}, qualitative analysis of comments to identify changes in online dialogue over time \cite{yardi_dynamic_2010}, and visualization research that reverse-engineers and classifies chart images  \cite{yardi_dynamic_2010, savva_revision_2011}.


\subsection{Twitter data and quantitative analysis} 

\subsubsection{Dataset} This analysis is conducted using a dataset of tweet IDs that Emily Chen et al. \cite{chen_public_2020} assembled by monitoring the Twitter streaming API for certain keywords associated with COVID-19 (e.g., ``coronavirus'', ``pandemic'', ``lockdown'', etc.) as well as by following the accounts of public health institutions (e.g., @CDCGov, @WHO, etc). We used version 2.7 of this dataset, which included over 390M tweets spanning January 21, 2020--July 31, 2020. This dataset consists of tweet IDs which we ``hydrated'' using twarc \cite{documenting_the_now_twarc_2016} into full tweets with associated metadata (e.g., hashtags, mentions, replies, favorites, retweets, etc.). 

To identify tweets that primarily discuss data visualizations about the pandemic, we initially adopted a strategy that filtered for tweets that contained at least one image and keyword associated with data analysis (e.g., ``bar,'' ``line,'' but also ``trend,'' ``data''). Unfortunately, this strategy yielded more noise than signal as most images in the resultant dataset were memes and photographs. We therefore adopted a more conservative approach, filtering for tweets that explicitly mentioned chart-related keywords (i.e., ``chart(s)'', ``plot(s)'', ``map(s)'', ``dashboard(s)'', ``vis'', ``viz'', or ``visualization(s)''). This process yielded a dataset of almost 500,000 tweets that incorporated over 41,000 images. We loaded the tweets and their associated metadata into a SQLite database, and the images were downloaded and stored on the file system.

\subsubsection{Image classification} To analyze the types of visualization found in the dataset, we began by classifying every image in our corpus using the mark classification computer vision model trained by Poco and Heer \cite{poco_reverseengineering_2017}. Unfortunately, this model was only able to classify 30\% of the images. As a result, we extracted a 4096-dimensional feature embedding for every image, and ran k-means clustering on a 100-dimensionally reduced space of these embeddings, for steps of k from 5--40. Two authors manually inspected the outputs of these runs, independently identified the most salient clusters, and then cross validated their analysis to assemble a final list of 8 relevant clusters: line charts, area charts, bar charts, pie charts, tables, maps, dashboards, and images. For dimensionality reduction and visualization, we used the UMAP algorithm \cite{mcinnes_umap_2018} and iteratively arrived at the following parameter settings: 20 neighbors, a minimum distance of 0.01, and using the cosine distance metric. To account for UMAP’s stochasticity, we executed 10 runs and qualitatively examined the output to ensure our analyses were not based on any resultant artifacts. 

\subsubsection{Network analysis} Finally, to analyze the users participating in these discussions, we constructed a network graph: nodes were users who appeared in our dataset, and edges were drawn between pairs of nodes if a user mentioned, replied to, retweeted, or quote-tweeted another user. The resultant graph consisted of almost 400,000 nodes and over 583,000 edges, with an average degree of 2.9. To produce a denser network structure, we calculated a histogram of node degrees and identified that two-thirds of the nodes were degree 1. We then computed subgraphs, filtering for nodes with a minimum degree of 2 through 10 and found that degree 5 offered us a good balance between focusing on the most influential actors in the network, without losing smaller yet salient communities. This step yielded a subgraph of over 28,000 nodes and 104,000 edges, with an average degree of 7.3. We detected communities on this network using the Louvain method \cite{blondel_fast_2008}. Of the 2,573 different communities detected by this algorithm, we primarily focus on the top 10 largest communities which account for 72\% of nodes, 80\% of edges, and 30\% of visualizations.


\subsection{Facebook data and qualitative analysis}

\subsubsection{Digital ethnography} While qualitative research can involve clinical protocols like interviews or surveys, Clifford Geertz \cite{geertz_deep_1998} argues that the most substantial ethnographic insights into the cultural life of a community come from ``\textbf{deep hanging out},'' i.e., long-term, participant observation alongside its members. Using ``lurking,'' a mode of participating by observing specific to digital platforms, we propose ``\textbf{deep lurking}'' as a way of systematically documenting the cultural practices of online communities. Our methods here rely on robust methodological literature in digital ethnography \cite{markham_fieldwork_2013,coleman_hacker_2014}, and we employ a case study approach \cite{small_how_2009} to analyze these Facebook groups. To that end, we followed five Facebook groups (each with a wide range of followers, 10K-300K) over the first six months of the coronavirus pandemic, and we collected posts throughout the platform that included terms for ``coronavirus'' and ``visualization'' with Facebook’s CrowdTangle tool~\cite{crowdtangle_team_crowdtangle_2020}. In our deep lurking, we archived web pages and took field notes on the following: posts (regardless of whether or not they included ``coronavirus'' and ``data''), subsequent comments, Facebook Live streams, and photos of in-person events. We collected and analyzed posts from these groups from their earliest date to September 2020. 

Taking a case study approach to the interactional Facebook data yields an analysis that ultimately complements the quantitative analysis. While the objective with analyzing Twitter data is statistical representativeness---we investigate which visualizations are the most popular, and in which communities---the objective of analyzing granular Facebook data is to accurately understand social dynamics within a singular community~\cite{small_how_2009}. As such, the Twitter and Facebook analyses are foils of one another: we have the ability to quantitatively analyze large-scale interactions on Twitter, whereas we analyze the Facebook data by close reading and attending to specific context. Twitter communities are loosely formed by users retweeting, liking, or mentioning one another; Facebook groups create clearly bounded relationships between specific communities. By matching the affordances of each data source with the most ecologically appropriate method (network analysis and digital ethnography), this paper meaningfully combines qualitative and quantitative methods to understand data visualizations about the pandemic on a deeply contextual level and at scale. 

\subsubsection{Data collection \& analysis} Concretely, we printed out posts as PDFs, tagged them with qualitative analysis software, and synthesized themes across these comments using grounded theory~\cite{kozinets_integrating_2019}. Grounded theory is an inductive method where researchers collect data and tag it by identifying analytically pertinent themes. Researchers then group these codes into higher-level concepts. As Kathy Charmaz \cite{charmaz_constructing_2006} writes: these ``methods consist of systematic, yet flexible guidelines for collecting and analyzing qualitative data to construct theories ‘grounded’ in the data themselves,'' and these methods have since been adapted for social media analysis~\cite{postill_social_2012}. While this flexibility allows this method to respond dynamically to changing empirical phenomena, it can also lead to ambiguity about how new data fit with previously identified patterns. Digital ethnography also requires a longer time horizon than quantitative work in order to generate meaningful insights and, on its own, does not lead immediately to quantifiable results. These limitations are a major reason to use both qualitative and quantitative approaches. Following Emerson et al. \cite{emerson_writing_2011}, we employ an integrative strategy that weaves together ``exemplars'' from qualitative data alongside our interpretations. We have redacted the names of individual users and the Facebook groups we have studied, but we have preserved the dates and other metadata of each post within the article where possible.

\subsubsection{Note on terminology} Throughout this study, we use the term ``anti-mask'' as a synecdoche for a broad spectrum of beliefs: that the pandemic is exaggerated, schools should be reopening, etc. While groups who hold these beliefs are certainly heterogeneous, the mask is a common flashpoint throughout the ethnographic data, and they use the term ``maskers'' to describe people who are driven by fear. They are ``anti-mask'' by juxtaposition. This study therefore takes an emic (i.e. ``insider'') approach to analyzing how members of these groups think, talk, and interact with one another, which starts by using terms that these community members would use to describe themselves. There is a temptation in studies of this nature to describe these groups as ``anti-science,'' but this would make it completely impossible for us to meaningfully investigate this article's central question: understanding what these groups mean when they say ``science.'' 

\section{Case Study} 

In the Twitter analysis, we quantitatively examine a corpus of tweets that use data visualizations to discuss the pandemic, and we create a UMAP visualization (figure 1) that identifies the types of visualizations that proliferate on Twitter. Then, we create a network graph (figure 2) of the users who share and interact with these data visualizations; the edges that link users in a network together are retweets, likes, mentions. We discover that the fourth largest network in our data consists of users promulgating heterodox scientific positions about the pandemic (i.e., anti-maskers). By comparing the visualizations shared within anti-mask and mainstream networks, we discover that there is no significant difference in the kinds of visualizations that the communities on Twitter are using to make drastically different arguments about coronavirus (figure 3). Anti-maskers (the community with the highest percentage of verified users) also share the second-highest number of charts across the top six communities (table 1), are the most prolific producers of area/line charts, and share the fewest number of photos (memes and images of politicians; see figure 3). Anti-maskers are also the most likely to amplify messages from their own community. We then examine the kinds of visualizations that anti-maskers discuss (figure 4). 

This leads us to an interpretive question that animates the Facebook analysis: how can opposing groups of people use similar methods of visualization and reach such different interpretations of the data? We approach this problem by ethnographically studying interactions within a community of anti-maskers on Facebook to better understand their practices of knowledge-making and data analysis, and we show how these discussions exemplify a fundamental epistemological rift about how knowledge about the coronavirus pandemic should be made, interpreted, and shared.


\subsection{Visualization design and network analysis} 

\begin{figure*}[htbp]
\centering
\includegraphics[width=0.90\textwidth]{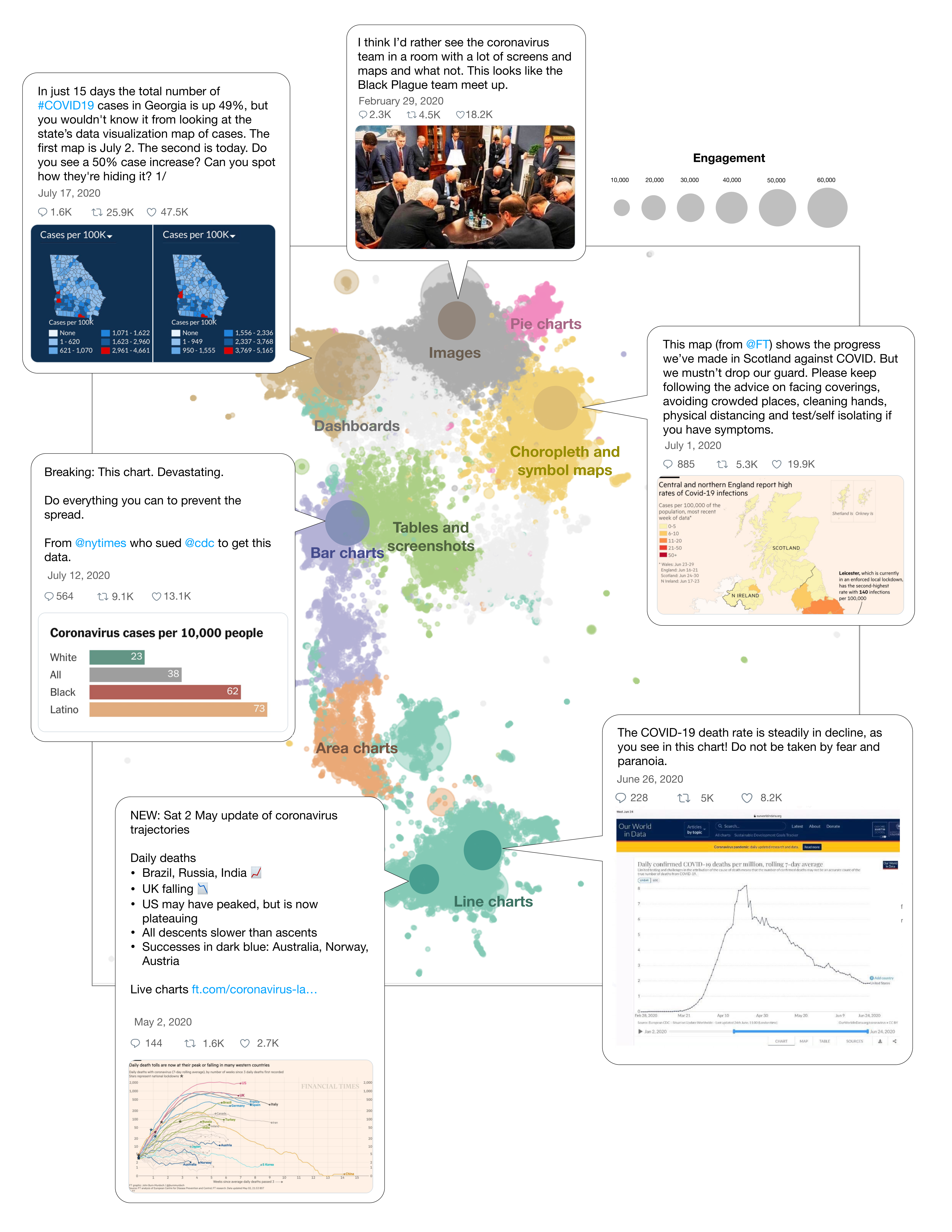}
  \caption{A UMAP visualization of feature embeddings of media found in our Twitter corpus. Color encodes labeled clusters, and size encodes the amount of engagement the media received (i.e., the sum of replies, favorites, retweets, and quote tweets).}
  \Description{Scatterplot of the different types of visualizations that are present within the corpus labeled with their types (dashboards, images, choropleth/symbol maps, tables/screenshots, bar charts, area charts, line charts). Each cluster has a callout with a tweet and attached visualization. For bar charts: tweet from July 12, 2020 which includes a bar chart about coronavirus cases per 10k people separated by race: ``Breaking: This chart. Devastating. Do everything you can to prevent the spread. From @nytimes who sued @cdc to get this data.'' For line charts, we have two tweets. The first line chart from May 2, 2020 includes a Financial Times line chart with multiple countries displaying that daily death tolls are at their peak or falling in many Western countries. The second line chart from June 26, 2020 is on daily confirmed COVID-19 deaths per million, rolling 7-day average: ``The COVID-19 death rate is steadily in decline, as you see in this chart! Do not be taken by fear and paranoia.`` For dashboards: tweet from July 17, 2020 with two choropleth maps of Georgia side-by-side, one from July 2 and the other from July 17 with cases per 100k. ``In just 15 days the total number of #COVID19 cases in Georgia is up 49\%, but you wouldn't know it from looking at the state’s data visualization map of cases.'' For images: a tweet from February 29, 2020 with an image of White House officials with their heads bowed and looking solemn in a room: ``I think I’d rather see the coronavirus team in a room with a lot of screens and maps and what not. This looks like the Black Plague team meet up.'' For choropleth and symbol maps: a tweet from July 1, 2020 with a map of central and northern England COVID-19 infection rates: ``This map (from @FT) shows the progress we’ve made in Scotland against COVID. But we mustn’t drop our guard.'' There are no callout tweets for the visualization types of area charts and tables/screenshots.}
\end{figure*}

\subsubsection{Visualization types} What kinds of visualizations are Twitter users sharing about the pandemic? Figure 1 visualizes the feature embeddings of images in our corpus, with color encoding clusters revealed and manually curated through k-means. Each circle is sized by the engagement the associated tweet received calculated as the sum of the number of favorites, replies, retweets, and quote tweets. Our analysis revealed eight major clusters: line charts (8908 visualizations, 21\% of the corpus), area charts (2212, 5\%), bar charts (3939, 9\%), pie charts (1120, 3\%), tables (4496, 11\%), maps (5182, 13\%), dashboards (2472, 6\%), and images (7,128, 17\%). The remaining 6,248 media (15\% of the corpus) did not cluster in thematically coherent ways. Here, we characterize salient elements and trends in these clusters.

\textbf{Line charts} represent the largest cluster of visualizations in our corpus. There are three major substructures: the first comprises line charts depicting the exponential growth of cases in the early stages of the pandemic, and predominantly use log-scales rather than linear scales. Charts from John Burn-Murdoch at the \textit{Financial Times} and charts from the nonprofit Our World in Data are particularly prominent here. A second substructure consists of line charts comparing cases in the United States and the European Union when the US was experiencing its second wave of cases, and the third consists of line charts that visualize economic information. This substructure includes line charts of housing prices, jobs and unemployment, and stock prices (the latter appear to be taken from financial applications and terminals, and often feature additional candlestick marks). Across this cluster, these charts typically depict national or supranational data, include multiple series, and very rarely feature legends or textual annotations (other than labels for each line). Where they do occur, it is to label every point along the lines. Features of the graph are visually highlighted by giving some lines a heavier weight or graying other ones out.

\textbf{Maps} are the second largest cluster of visualizations in our corpus. The overwhelming majority of charts here are choropleths (shaded maps where a geographic region with high COVID rates might be darker, while low-rate regions are lighter). Other visualizations in this cluster include cartograms (the size of a geographic region is proportional its number of COVID infections as a method of comparison) and symbol maps (the size of a circle placed on a geographic region is proportional to COVID infections). 
The data for these charts span several geographic scales---global trends, country-level data (the US, China, and the UK being particularly salient), and municipal data (states and counties). These maps generally feature heavy annotation including direct labeling of geographic regions with the name and associated data value; arrows and callout boxes also better contextualize the data. For instance, in a widely shared map of the United Kingdom from the \textit{Financial Times}, annotations described how “[t]hree Welsh areas had outbreaks in meatpacking plants in June” and that “Leicester, which is currently in an enforced local lockdown, has the second-highest rate...” These maps depict a wide range of data values including numbers of cases/deaths, metrics normalized per capita, rate of change for cases and/or deaths, mask adherence rates, and the effect of the pandemic on greenhouse gas emissions. Interestingly, choropleth maps of the United States electoral college at both the state- and district-level also appear in the corpus, with the associated tweets comparing the winner of particular regions with the type of pandemic response.

\textbf{Area charts} feature much heavier annotation than line charts (though fewer than maps). Peaks, troughs, and key events (e.g., when lockdowns occurred or when states reopened) are often shaded or labeled with arrows, and line marks are layered to highlight the overall trend or depict the rolling average. When these charts reflect data with a geographic correspondence, this data is often at a more local scale; line charts typically depict national or supranational data, and area charts more often visualized data at the state or county level. Notable subclusters in this group include the viral ``Flatten the Curve'' graphic, stacked area/streamgraphs, and ``skinny bar'' charts (charts of temporal data that closely resemble area charts, but use bar marks with narrow widths. Charts from the \textit{New York Times} are especially prominent examples of the latter category---particularly screenshots of a red chart that was featured on the mobile front page.

\textbf{Bar charts} are predominantly encode categorical data and are more consistently and more heavily annotated than area charts. In addition to the annotations described for area charts (direct labeling of the tops of bars, labeled lines and arrows), charts in this cluster often include concise explainer texts. 

These texts include some form of extended subtitles, more descriptive axis tick labels, or short passages before or after the bar chart that contextualize the data. Visually, the cluster is equally split between horizontal and vertical charts, and both styles feature a mix of layered, grouped, and stacked bars. Bar chart ``races'' (e.g., those developed with the Flourish visualization package) are one of the more frequently recurring idioms in this cluster. These are horizontal bar charts depicting the total number of cases per country, and animated over time.

\textbf{Dashboards and images.} While the remaining clusters are thematically coherent, we did not observe as rich a substructure within them. The dashboard cluster is overwhelmingly dominated by screenshots of the Johns Hopkins dashboard, and the image cluster is primarily comprised of reaction memes featuring the photos or caricatures of heads of state.

\begin{figure}[t]
  \centering
  \includegraphics[width=\linewidth]{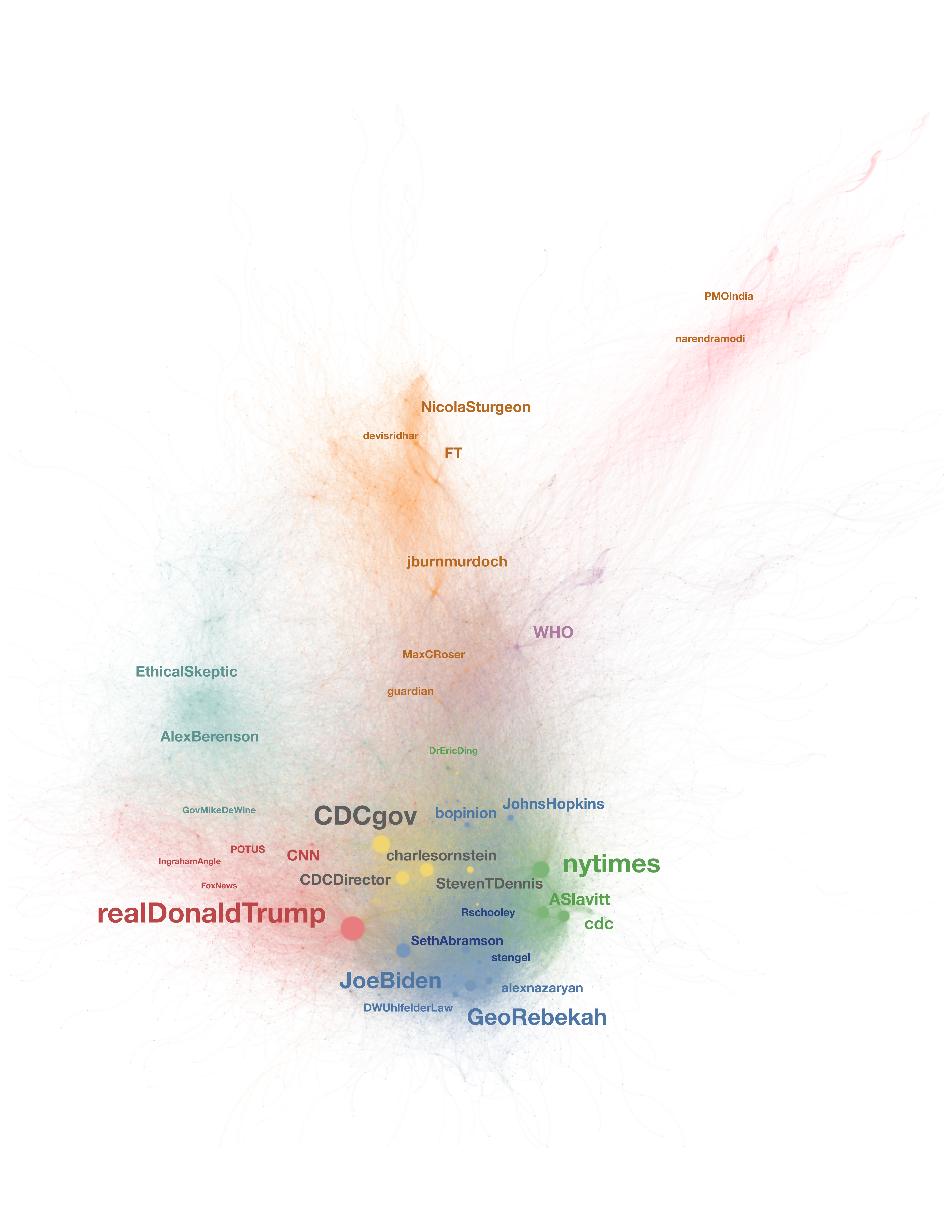}
  \caption{A network visualization of Twitter users appearing in our corpus. Color encodes community as detected by the Louvain method \cite{blondel_fast_2008}, and nodes are sized by their degree of connectedness (i.e., the number of other users they are connected to). }
  \Description{A network visualization of the Twitter corpus, with colorings for clusters and labels for the most prominent Twitter users. The teal cluster has users @EthicalSkeptic and @AlexBerenson. The orange cluster has @POTUS, @CNN, and @realDonaldTrump. The blue cluster has @JoeBiden, @GeoRebekah, @JohnsHopkins, @bopinion, @Rshooley, @SethAbramson, @stengel, and @alexnazaryan. The green cluster has @nytimes, @ASlavitt, @cdc, and @DrEricDing. The red cluster has @FT, @jburnmurdoch,@ NicolaSturgeon, @devisridhar, @MaxCRoser, @guardian. The purple cluster has @WHO.}
\end{figure}

\begin{table*}[h]
  \caption{Descriptive Statistics of Communities}
  \label{tab:freq}
  \begin{tabular}{c c c c c}
    \toprule
    Community \# & Verified Users as \% of Total Users  & In-Network Retweets as \% of Total Retweets   & Original Tweets as \% of Total Tweets \\
    \midrule
  1 & 8.39      & 73.30     & 22.12 \\
  2 & 14.36     & 75.45     & 44.75 \\
  3 & 22.92     & 89.32     & 34.00 \\
  4 & 10.56     & 82.17     & 37.12 \\
  5 & 12.33     & 58.29     & 21.57 \\
  6 & 8.94      & 70.97     & 37.46 \\
  \bottomrule
\end{tabular}
\end{table*}

\begin{figure*}[h]
  \centering
  \includegraphics[width=\textwidth]{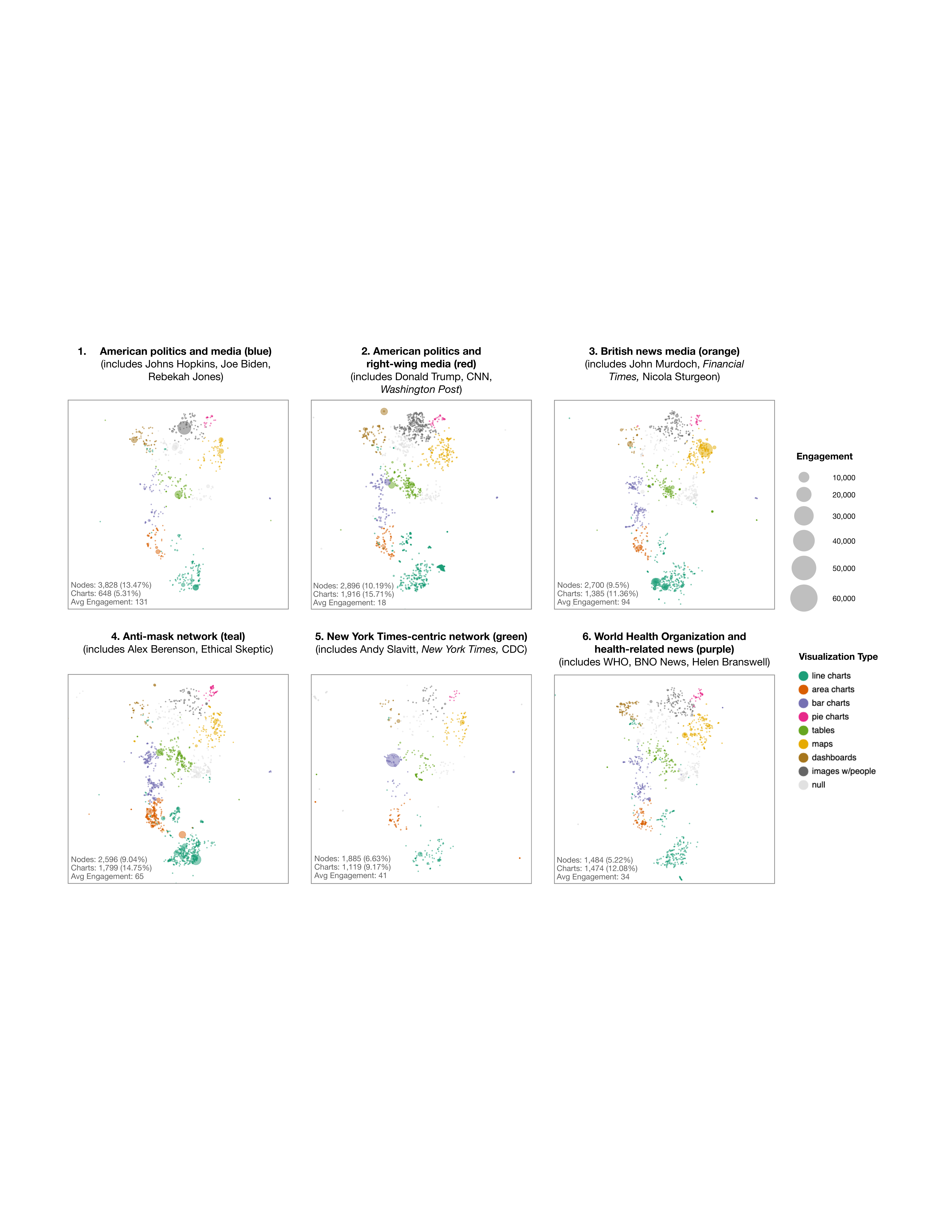}
  \caption{Visualizing the distribution of chart types by network community (with top accounts listed). While every community has produced at least one viral tweet, anti-mask users (group 6) receive higher engagement on average. }
  \Description{A display of six scatter plots, each one representing a specific cluster of users and the visualizations that they employed. There are two rows and three columns. For the first row: the first one has a title of American politics and media (blue) with highlighted users Johns Hopkins, Joe Biden, Rebekah Jones. In the bottom left, there are metrics of the network. Nodes: 3,828 (13.47\%), charts: 648 (5.31\%), average engagement: 131. The second one has a title of American politics and right-wing media (red) with highlighted users Donald Trump, CNN, and Washington Post. In the bottom left, there are metrics of the network. Nodes: 2,896 (10.19\%), charts: 1,916 (15.17\%), average engagement: 18. The third one has a title of British news media (orange) with highlighted users John Murdoch, Financial Times, and Nicola Sturgeon. In the bottom left, there are metrics of the network. Nodes: 277 (9.5\%), charts: 1,385 (11.36\%), average engagement: 94. For the second row: the fourth one has a title of Anti-mask network (teal) with highlighted users Alex Berenson and Ethical Skeptic. In the bottom left, there are metrics of the network. Nodes: 2,596 (9.04\%), charts: 1,799 (14.75\%), average engagement: 65. The fifth one has a title of New York Times-centric network (green) with highlighted users Andy Slavitt, New York Times, and CDC. In the bottom left, there are metrics of the network. Nodes: 1,885 (6.63\%), charts: 1,119 (9.17\%), average engagement: 41. The sixth one has a title of World Health Organization and health-related news (purple) with highlighted users WHO, BNO News, and Helen Branswell. In the bottom left, there are metrics of the network. Nodes: 1,484 (5.22\%), charts: 1,474 (12.08\%), average engagement: 34. To the right of all of the scatter plots, there’s a legend indicating engagement sizes and visualization colors.}
\end{figure*}

\begin{figure*}[hbtp]
  \centering
  \includegraphics[width=0.85\linewidth]{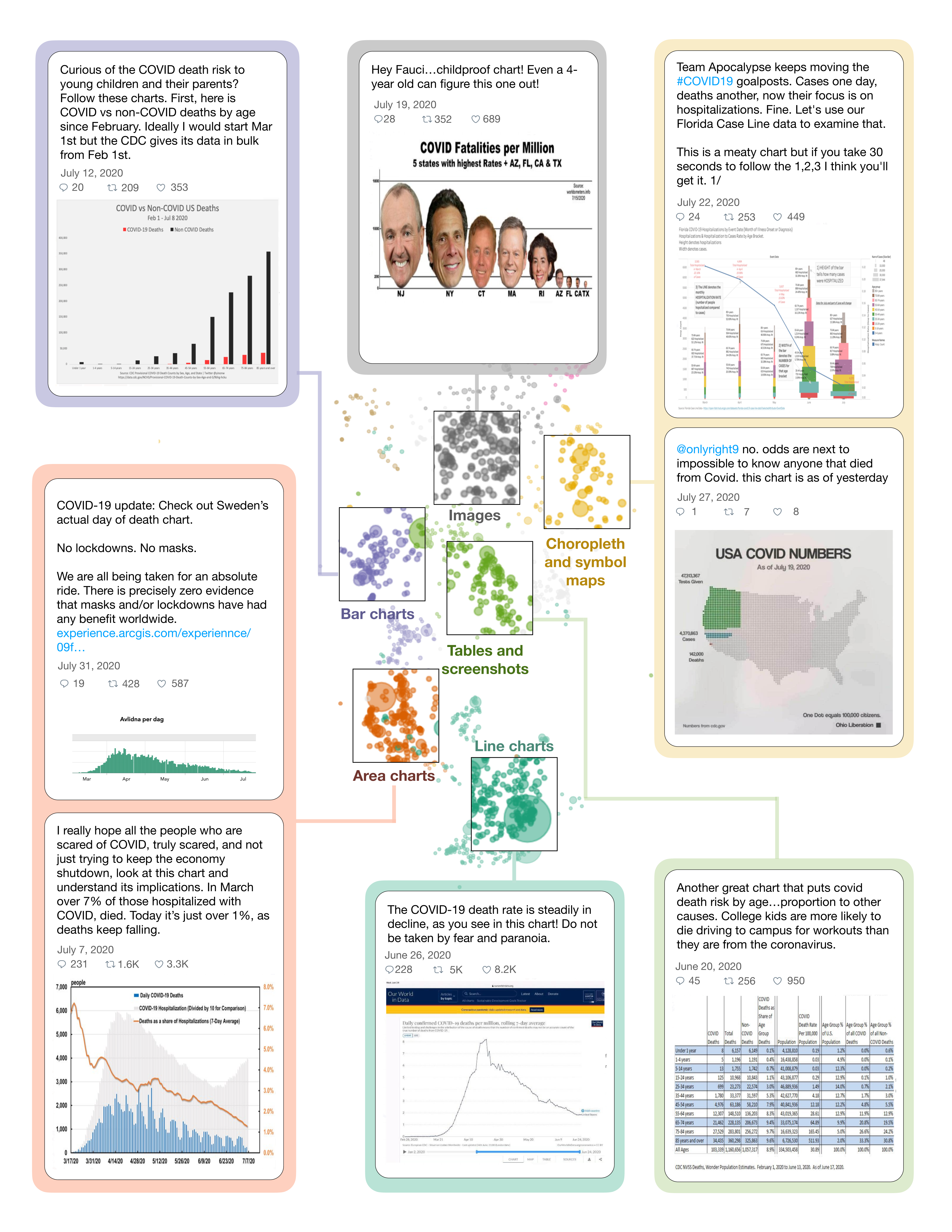}
  \caption{Sample counter-visualizations from the anti-mask user network. While there are meme-based visualizations, anti-maskers on Twitter adopt the same visual vocabulary as visualization experts and the mainstream media.}
  \Description{Scatterplot of the different types of visualization within the anti-mask network, each labeled by type (area charts, bar charts, images, line charts, table/screenshots, choropleth/symbol maps). Each cluster has a callout of a tweet and attached visualization. For bar chart: tweet from July 12, 2020 that compares COVID-19 deaths and non COVID deaths by age. For area charts, we have two tweets. The first tweet is from July 7, 2020 and displays daily COVID-19 deaths, COVID-19 hospitalizations, and deaths. The second tweet from July 31, 2020 has a chart of deaths per day from March to July: ``COVID-19 update: Check out Sweden’s actual day of death chart. No lockdowns. No masks. We are all being taken for an absolute ride. There is precisely zero evidence that masks and/or lockdowns have had any benefit worldwide.'' For images: tweet from July 19, 2020 with chart displaying COVID fatalities per million per state uses the faces of the governors of the respective states and their relative scaled sizes to emphasize fatalities. For choropleth/symbol maps, there are two tweets. The first one from July 22, 2020 displays column charts throughout the months of quarantine. The second tweet from July 27, 2020 has a chart of USA COVID numbers as of July 19, 2020 with a highlight on the Pacific Northwest. For tables/screenshots: tweet from June 20, 2020 with a table breaking down COVID-related deaths and total deaths: ``Another great chart that puts covid death risk by age…proportion to other causes. College kids are more likely to die driving to campus for workouts than they are from the coronavirus.'' For line chart: tweet from June 26, 2020 with a screenshot of a line graph of daily confirmed COVID-19 deaths per million, rolling 7-day average: ``The COVID-19 death rate is steadily in decline, as you see in this chart! Do not be taken by fear and paranoia.''}
\end{figure*}

\subsubsection{User networks} 

What are the different networks of Twitter users who share COVID-related data visualizations, and how do they interact with one another? Figure 2 depicts a network graph of Twitter users who discuss (or are discussed) in conversation with the visualizations in Figure 1. This network graph only shows users who are connected to at least five other users (i.e., by replying to them, mentioning them in a tweet, or re-tweeting or quote-tweeting them). The color of each network encodes a specific community as detected by the Louvain method \cite{blondel_fast_2008}, and the graph accounts for the top 10 communities (20,500 users or 72\% of the overall graph). We describe the top six networks below listed in order of size (i.e., number of users within each network). While we have designated many of these communities with political orientation (e.g., left- or right-wing), these are only approximations; we recognize that these terms are fundamentally fluid and use them primarily as shorthand to make cross-network comparisons (e.g., mainstream political/media organizations vs. anti-mask protestors). 

\textbf{1. American politics and media (blue).} This community features the American center-left, left, mainstream media, and popular or high profile figures (inside and outside of the scientific community). Accounts include politicians (@JoeBiden, @SenWarren), reporters (@joshtpm, @stengel), and public figures like Floridian data scientist @GeoRebekah and actor @GeorgeTakei. The user with the most followers in this community is @neiltyson. 

\textbf{2. American politics and right-wing media (red).} This community includes members of the Trump administration, Congress, and right-wing personalities (e.g., @TuckerCarlson). Several accounts of mainstream media organizations also lie within this community (@CNN, @NBCNews), which reflects how often they mention the President (and other government accounts) in their coverage. Notably, these are official organizational accounts rather than those of individual reporters (which mostly show up in the previous group). Several mainstream media organizations are placed equally between these two clusters (@NPR, @washingtonpost). The user with the most followers in this community is @BarackObama. 

\textbf{3. British news media (orange)}. The largest non-Americentric network roughly corresponds to news media in the UK, with a significant proportion of engagement targeted at the \emph{Financial Times'} successful visualizations by reporter John Burn-Murdoch, as well as coverage of politician Nicola Sturgeon's coronavirus policies. The user with the most followers  in this community is @TheEconomist. 

\textbf{4. Anti-mask network (teal).} The anti-mask network comprises over 2,500 users (9\% of our network graph) and is anchored by former \emph{New York Times} reporter @AlexBerenson, blogger @EthicalSkeptic, and @justin\_hart. \emph{The Atlantic}’s @Covid19Tracking project (which collates COVID-19 testing rates and patient outcomes across the United States) and @GovMikeDeWine are also classified as part of this community. Governor DeWine of Ohio is not an anti-masker, but is often the target of anti-mask protest given his public health policies. Anti-mask users also lampoon \emph{The Atlantic}'s project as another example of mainstream misinformation. These dynamics of intertextuality and citation within these networks are especially important here, as anti-mask groups often post screenshots of graphs from ``lamestream media'' organizations (e.g., \emph{New York Times}) for the purpose of critique and analysis. The user with the most followers in this community is COVID skeptic and billionaire @elonmusk.

\textbf{5. New York Times-centric network (green).} This community is largely an artifact of a single visualization: Andy Slavitt (@ASlavitt), the former acting Administrator of the Centers for Medicare and Medicaid Services under the Obama administration, posted a viral tweet announcing the \emph{New York Times} had sued the CDC (tagged with the incorrect handle @cdc instead of @CDCGov). The attached bar chart showing the racial disparity in COVID cases was shared widely with commentary directly annotated onto the graph itself, or users analyzed the graph through quote-tweets and comments. The user with the most followers in this community is @NYTimes. 

\textbf{6. World Health Organization and health-related news organizations (purple).} This community consists of global health organizations, particularly the @WHO and its subsidiary accounts (e.g., @WHOSEARO for Southeast Asia). The user with the most followers in this community is @YouTube.

\subsubsection{Descriptive statistics of communities} 
Table 1 lists summary statistics for the six largest communities in our dataset. There are three statistics of interest: the percentage of verified users (based on the total number of users within a community), the percentage of in-network retweets (based on a community's total number of retweets), and the percentage of original tweets (based on a community's total number of tweets). Twitter verification can often indicate that the account of public interest is authentic (subject to Twitter's black-boxed evaluation standards); it can be a reasonable indication that the account is not a bot. Secondly, a high percentage of in-network retweets can be an indicator of how insular a particular network can be, as it shows how often a user amplifies messages from other in-network members. Finally, the percentage of original tweets shows how much of the content in that particular community is organic (i.e., they write their own content rather than simply amplifying existing work). Communities that have users who use the platform more passively (i.e., they prefer to lurk rather than comment) will have fewer original tweets; communities that have higher levels of active participation will have a higher number of original tweets as a percentage of total tweets.

The networks with the highest number of in-network retweets (which can be one proxy for insularity) are the British media (89.32\%) and the anti-mask networks (82.17\%), and the network with the highest percentage of original tweets is the American politics and right-wing media network (44.75\%). Notably, the British news media network has both the \textit{highest} percentage of verified users (22.92\%), the \textit{highest} percentage of in-network retweets (89.32\%), and the fourth-highest percentage of original tweets (34.00\%). As the third largest community in our dataset, we attribute this largely to the popularity of the graphs from the \textit{Financial Times} from a few sources and the constellation of accounts that discussed those visualizations. While other communities (anti-mask, American politics/right-wing media, and WHO/health-related news) shared \emph{more} visualizations, this network shared fewer graphs (1,385) that showed the \emph{second-highest} level of engagement across the six communities (averaging 94 likes, retweets, or mentions per visualization). The network whose visualizations garner the highest level of engagement is the American politics and media network (131 likes, retweets, or mentions per visualization), but they only shared about half of the visualizations (648) compared to their British counterparts.

Through the descriptive statistics, we find that the anti-mask community exhibits very similar patterns to the rest of the networks in our dataset (it has about the same number of users with the same proportion of verified accounts). However, this community has the second highest percentage of in-network retweets (82.17\% of all retweets) across the communities, and has the third-highest percentage of original tweets (37.12\%, only trailing the World Health Organization network, at 37.46\%). 

\subsubsection{Cross-network comparison of visualization types} 
Figure 3 depicts the distribution of visualization types by each community, along with descriptive statistics on the numbers of users, charts, and average engagement per tweet. These scatterplots show that there is little variance between the types of visualizations that users in each network share: almost all groups equally use maps or line, area, and bar charts. However, each group usually has one viral visualization---in group 3 (British media), the large yellow circle represents a map from the \emph{Financial Times} describing COVID-19 infections in Scotland; in group 5 (\emph{New York Times}), the large purple circle in the center of the chart represents the viral bar chart from Andy Slavitt describing the racial disparities in COVID cases. The visualizations with the highest number of engagements in each of the six communities is depicted in figure 1. 

Overall, we see that each group usually has one viral hit, but that the anti-mask users (group 4) tend to share a wide range of visualizations that garner medium levels of engagement (they have the third-highest number of average engagements in the six communities; an average of 65 likes, shares, and retweets). As a percentage of total tweets, anti-maskers have shared the second-highest number of charts across the top six communities (1,799 charts or 14.75\%). They also use the \emph{most} area/line charts and the \emph{least} images across the six communities (images in this dataset usually include memes or photos of politicians). These statistics suggest that anti-maskers tend to be among the most prolific sharers of data visualizations on Twitter, and that they overwhelmingly amplify these visualizations to other users within their network (88.97\% of all retweets are in-network). 

\subsubsection{Anti-mask visualizations} 
Figure 4 depicts the data visualizations that are shared by members of the anti-mask network accompanied by a select tweets from each category. While there are certainly visualizations that tend to use a meme-based approach to make their point (e.g., ``Hey Fauci...childproof chart!'' with the heads of governors used to show the rate of COVID fatalities), many of the visualizations shared by anti-mask Twitter users employ visual forms that are relatively similar to charts that one might encounter at a scientific conference. Many of these tweets use area and line charts to show the discrepancy between the number of projected deaths in previous epidemiological and the numbers of actual fatalities. Others use unit visualizations, tables, and bar charts to compare the severity of coronavirus to the flu. In total, this figure shows the breadth of visualization types that anti-mask users employ to illustrate that the pandemic is exaggerated. 


\subsection{Anti-mask discourse analysis} 

The Twitter analysis establishes that anti-maskers are prolific producers and consumers of data visualizations, and that the graphs that they employ are similar to those found in orthodox narratives about the pandemic. Put differently, anti-maskers use ``data-driven'' narratives to justify their heterodox beliefs. However, a quantitative overview of the visualizations they share and amplify does not in itself help us understand \emph{how} anti-maskers invoke data and scientific reasoning to support policies like re-opening schools and businesses. Anti-maskers are acutely aware that mainstream narratives use data to underscore the pandemic's urgency; they believe that these data sources and visualizations are fundamentally flawed and seek to counteract these biases. This section showcases the different ways that anti-mask groups talk about COVID-related data in discussion forums. What kinds of concerns do they have about the data used to formulate public policies? How do they talk about the limitations of data or create visualizations to convince other members in their physical communities that the pandemic is a hoax? 


\subsubsection{Emphasis on original content} 

Many anti-mask users express mistrust for academic and journalistic accounts of the pandemic, proposing to rectify alleged bias by ``following the data'' and creating their own data visualizations. Indeed, one Facebook group within this study has very strict moderation guidelines that prohibit the sharing of non-original content so that discussions can be ``guided solely by the data.'' Some group administrators even impose news consumption bans on themselves so that ``mainstream'' models do not ``cloud their analysis.'' In other words, \textbf{anti-maskers value unmediated access to information and privilege personal research and direct reading over ``expert'' interpretations}. While outside content is generally prohibited, Facebook group moderators encourage followers to make their own graphs, which are often shared by prominent members of the group to larger audiences (e.g., on their personal timelines or on other public facing Pages). Particularly in cases where a group or page is led by a few prominent users, follower-generated graphs tend to be highly popular because they often encourage other followers to begin their own data analysis projects, and comments on these posts often deal directly with how to reverse-engineer (or otherwise adjust) the visualization for another locality. 

Since some of these groups are focused on a single state (e.g. ``Reopen Nevada''), they can fill an information gap: not every county or locality is represented on data dashboards made by local newspapers, health departments, or city governments---if these government entities have dashboards or open data portals at all. In such cases, the emphasis on original content primarily reflects a grassroots effort to ensure access to pandemic-related data where there are no alternatives, and only secondarily serves to constitute an alternative to ideologically charged mainstream narratives. In the rare instances where mainstream visualizations are shared in such a group, it is usually to highlight the ways that mainstream analysis finally matches anti-mask projections, or to show how a journalist, government official, or academic can manipulate the same data source to purposefully mislead readers.

In order to create these original visualizations, users provide numerous tutorials on how to access government health data. These tutorials come either as written posts or as live screencasts, where a user (often a group administrator or moderator) demonstrates the process of downloading information from an open data portal. During the livestream, they editorialize to show which data are the most useful (e.g., \textit{``the data you can download} [from the Georgia Health Department website] \textit{is completely worthless, but the dashboard---which has data that everyday citizens cannot access---actually shows that there are no deaths whatsoever,''} July 13, 2020). Since many local health departments do not have the resources to stand up a new data system specifically for COVID-19, some redirect constituents to the state health department, which may not have granular data for a specific township available for public use. In the absence of data-sharing between states and local governments, users often take it upon themselves to share data with one another (e.g., ``[redacted] \textit{brings us this set of data from Minnesota.} [...] \textit{Here it is in raw form, just superimposed on the model,''} May 17, 2020) and they work together to troubleshoot problems with each dataset (e.g., \textit{``thanks. plugging} [sic] \textit{in new .csv file to death dates is frustrating but worth it,''} May 2, 2020). 


\subsubsection{Critically assessing data sources} 

Even as these users learn from each other to collect more data, they remain critical about the circumstances under which the data are collected and distributed. Many of the users believe that the most important metrics are missing from government-released data. They express their concerns in four major ways. First, there is an ongoing animated debate within these groups about which metrics matter. Some users contend that deaths, not cases, should be the ultimate arbiter in policy decisions, since case rates are easily ``manipulated'' (e.g., with increased testing) and do not necessarily signal severe health problems (people can be asymptomatic). The shift in focus is important, as these groups believe that the emphasis on cases and testing often means that rates of COVID deaths by county or township are not reported to the same extent or seriously used for policy making. As one user noted, \textit{``The Alabama public health department doesn’t provide deaths per day data (that I can tell---you can get it elsewhere). I sent a message asking about that. Crickets so far,''} (July 13, 2020). 

Second, users also believe that state and local governments are deliberately withholding data so that they can unilaterally make decisions about whether or not lockdowns are effective. During a Facebook livestream with a Congressional candidate who wanted to ``use data for reopening,'' for example, both the candidate and an anti-mask group administrator discussed the extent to which state executives were willing to obscure the underlying data that were used to justify lockdown procedures (August 30, 2020). To illustrate this, the candidate emphasized a press conference in which journalists asked the state executive whether they would consider making the entire contact tracing process public, which would include releasing the name of the bar where the outbreak started. In response, the governor argued that while transparency about the numbers were important, the state would not release the name of the bar, citing the possibility of stigmatization and an erosion of privacy. This soundbite---\textit{``we have the data, but we won’t give it to you''}---later became a rallying cry for anti-mask groups in this state following this livestream. \textit{``I hate that they’re not being transparent in their numbers and information they’re giving out,''} another user wrote. \textit{``They need to be honest and admit they messed up if it isn’t as bad as they’re making it out to be.} [...] \textit{We need honesty and transparency.''}

This plays into a third problem that users identify with the existing data: that datasets are constructed in fundamentally subjective ways. They are coded, cleaned, and aggregated either by government data analysts with nefarious intentions or by organizations who may not have the resources to provide extensive documentation. \textit{``Researchers can define their data set anyhow} [sic] \textit{they like in absence of generally accepted (preferably specified) definitions,''} one user wrote on June 23, 2020. \textit{``Coding data is a big deal---and those definitions should be offered transparently by every state. Without a national guideline---we are left with this mess.''} The lack of transparency within these data collection systems---which many of these users infer as a lack of honesty---erodes these users’ trust within both government institutions and the datasets they release. 

Even when local governments do provide data, however, these users also contend that the data requires context in order for any interpretation to be meaningful. For example, metrics like hospitalization and infection rates are still \textit{``vulnerable to all sorts of issues that make [these] data less reliable than deaths data''} (June 23, 2020), and require additional inquiry before the user considers making a visualization. In fact, there are multiple threads every week where users debate how representative the data are of the population given the increased rate of testing across many states. For some users, random sampling is the only way to really know the true infection rate, as (1) testing only those who show symptoms gives us an artificially high infection rate, and (2) testing asymptomatic people tells us what we already know---that the virus is not a threat. These groups argue that the conflation of asymptomatic and symptomatic cases therefore makes it difficult for anyone to actually determine the severity of the pandemic. \textit{``We are counting ‘cases’ in ways we never did for any other virus,''} a user writes, \textit{``and we changed how we counted in the middle of the game. It’s classic garbage in, garbage out at this point. If it could be clawed back to ONLY symptomatic and/or contacts, it could be a useful guide} [for comparison], \textit{but I don’t see that happening''} (August 1, 2020). 

Similarly, these groups often question the context behind measures like ``excess deaths.'' While the CDC has provided visualizations that estimate the number  excess deaths by week \cite{cdc_coronavirus_2020}, users take screenshots of the websites and debate whether or not they can be attributed to the coronavirus. \textit{``You can’t simply subtract the current death tally from the typical value for this time of year and attribute the difference to Covid,''} a user wrote. \textit{``Because of the actions of our governments, we are actually causing excess deaths. Want to kill an old person quickly? Take away their human interaction and contact. Or force them into a rest home with other infected people. Want people to die from preventable diseases? Scare them away from the hospitals, and encourage them to postpone their medical screenings, checkups, and treatments} [...] \textit{The numbers are clear. By trying to mitigate one problem, we are creating too many others, at too high a price''} (September 5, 2020). 


\subsubsection{Critically assessing data representations} 

Even beyond downloading datasets from local health departments, users in these groups are especially attuned to the ways that specific types of visualizations can obscure or highlight information. In response to a visualization where the original poster (OP) created a bar chart of death counts by county, a user commented: \textit{``the way data is presented can also show bias. For example in the state charts, counties with hugely different populations can be next to each other. The smaller counties are always going to look calm even if per capita they are doing the same or worse. Perhaps you could do a version of the charts where the hardest hit county is normalized per capita to 1 and compare counties that way,''} to which the OP responded, \textit{``it is never biased to show data in its entirety, full scale''} (August 14, 2020). 

An ongoing topic of discussion is whether to visualize absolute death counts as opposed to deaths per capita, and it is illustrative of a broader mistrust of mediation. For some, ``raw data'' (e.g., counts) provides more accurate information than any data transformation (e.g., death rate per capita, or even visualizations themselves). For others, screenshots of tables are the most faithful way to represent the data, so that people can see and interpret it for themselves. \textit{``No official graphs,''} said one user. \textit{``Raw data only. Why give them an opportunity to spin?''} (June 14, 2020). These users want to understand and analyze the information for themselves, free from biased, external intervention. 


\subsubsection{Identifying bias and politics in data} 

While users contend that their data visualizations objectively illustrate how the pandemic is no worse than the flu, they are similarly mindful to note that these analyses only represent partial perspectives that are subject to individual context and interpretation. \textit{``I’ve never claimed to have no bias. Of course I am biased, I’m human,''} says one prolific producer of anti-mask data visualizations. \textit{``That’s why scientists use controls... to protect ourselves from our own biases. And this is one of the reasons why I disclose my biases to you. That way you can evaluate my conclusions in context. Hopefully, by staying close to the data, we keep the effect of bias to a minimum''} (August 14, 2020). They are ultimately mindful of the subjectivity of human interpretation, which leads them to analyzing the data for themselves. 

More tangibly, however, these groups seek to identify bias by being critical about specific profit motives that come from releasing (or suppressing) specific kinds of information. Many of the users within these groups are skeptical about the potential benefits of a coronavirus vaccine, and as a point of comparison, they often reference how the tobacco industry has historically manipulated science to mislead consumers. These groups believe that pharmaceutical companies have similarly villainous profit motives, which leads the industry to inflate data about the pandemic in order to stoke demand for a vaccine. As one user lamented, \textit{``I wish more of the public would do some research into them and see how much of a risk they are but sadly most wont} [sic]---\textit{because once you do and you see the truth on them, you get labeled as an `antivaxxer' which equates to fool. In the next few years, the vaccine industry is set to be a nearly 105 billion dollar industry. People should really consider who profits off of our ignorance''} (August 24, 2020).


\subsubsection{Appeals to scientific authority}

Paradoxically, these groups also seek ways to validate their findings through the scientific establishment. Many users prominently display their scientific credentials (e.g., referring to their doctoral degrees or prominent publications in venues like Nature) which uniquely qualify them as insiders who are most well-equipped to criticize the scientific community. Members who perform this kind of expertise often point to 2013 Nobel Laureate Michael Levitt’s assertion that lockdowns do nothing to save lives \cite{lloyd_q_2020} as another indicator of scientific legitimacy. Both Levitt and these anti-mask groups identify the dangerous convergence of science and politics as one of the main barriers to a more reasonable and successful pandemic response, and they construct their own data visualizations as a way to combat what they see as health misinformation. \textit{``To be clear. I am not downplaying the COVID epidemic,''} said one user. \textit{``I have never denied it was real. Instead, I’ve been modeling it since it began in Wuhan, then in Europe, etc.} [...] \textit{What I have done is follow the data. I’ve learned that governments, that work for us, are too often deliberately less than transparent when it comes to reporting about the epidemic''} (July 17, 2020). For these anti-mask users, their approach to the pandemic is grounded in a \emph{more} scientific rigor, not less.


\subsubsection{Developing expertise and processes of critical engagement} 

The goal of many of these groups is ultimately to develop a network of well-informed citizens engaged in analyzing data in order to make measured decisions during a global pandemic. \textit{``The other side says that they use evidence-based medicine to make decisions,''} one user wrote, \textit{``but the data and the science do not support current actions''} (August 30, 2020). The discussion-based nature of these Facebook groups also give these followers a space to learn and adapt from others, and to develop processes of critical engagement. Long-time followers of the group often give small tutorials to new users on how to read and interpret specific visualizations, and users often give each other constructive feedback on how to adjust their graphic to make it more legible or intuitive. Some questions and comments would not be out of place at all at a visualization research poster session: \textit{``This doesn't make sense. What do the colors mean? How does this demonstrate any useful information?''} (July 21, 2020) These communities use data analysis as a way to socialize and enculturate their users; they promulgate data literacy practices as a way of inculcating heterodox ideology. The transmission of data literacy, then, becomes a method of political radicalization. 

These individuals as a whole are extremely willing to help others who have trouble interpreting graphs with multiple forms of clarification: by helping people find the original sources so that they can replicate the analysis themselves, by referencing other reputable studies that come to the same conclusions, by reminding others to remain vigilant about the limitations of the data, and by answering questions about the implications of a specific graph. The last point is especially salient, as it surfaces both what these groups see as a reliable measure of how the pandemic is unfolding and what they believe they should do with the data. These online communities therefore act as a sounding board for thinking about how best to effectively mobilize the data towards more measured policies like slowly reopening schools. \textit{``You can tell which places are actually having flare-ups and which ones aren’t,''} one user writes. \textit{``Data makes us calm.''} (July 21, 2020) 

Additionally, \textbf{followers in these groups also use data analysis as a way of bolstering social unity and creating a community of practice.} While these groups highly value scientific expertise, they also see collective analysis of data as a way to bring communities together within a time of crisis, and being able to transparently and dispassionately analyze the data is crucial for democratic governance. In fact, the explicit motivation for many of these followers is to find information so that they can make the best decisions for their families---and by extension, for the communities around them.  \textit{``Regardless of your political party, it is incumbent on all of us to ask our elected officials for the data they use to make decisions,''} one user said during a live streamed discussion. \textit{``I’m speaking to you as a neighbor: request the data.} [...] \textit{As a Mama Bear, I don’t care if Trump says that it’s okay, I want to make a decision that protects my kids the most. This data is especially important for the moms and dads who are concerned about their babies''} (August 30, 2020). As Kate Starbird et al. have demonstrated, strategic information operations require the participation of online communities to consolidate and amplify these messages: these messages become powerful when emergent, organic crowds (rather than hired trolls and bots) iteratively contribute to a larger community with shared values and epistemologies~\cite{starbird_disinformation_2019}.

Group members repost these analyses onto their personal timelines to start conversations with friends and family in hopes that they might be able to congregate in person. However, many of these conversations result in frustration. \textit{``I posted virus data from the CDC, got into discussion with people and in the end several straight out voiced they had no interest in the data,''} one user sighed. \textit{``My post said ‘Just the facts.’} [screenshot from the CDC] \textit{People are emotionally invested in their beliefs and won’t be swayed by data. It’s disturbing''} (August 14, 2020). Especially when these conversations go poorly, followers solicit advice from each other about how to move forward when their children’s schools close or when family members do not ``follow the data.'' One group even organized an unmasked get-together at a local restaurant where they passed out t-shirts promoting their Facebook group, took selfies, and discussed a lawsuit that sought to remove their state’s emergency health order (September 12, 2020). The lunch was organized such that the members who wanted to first attend a Trump rolling rally could do so and \textit{``drop in afterward for some yummy food and fellowship''} (September 8, 2020).


\subsubsection{Applying data to real-world situations} 

Ultimately, anti-mask users emphasize that they need to apply this data to real-world situations. The same group that organized the get-together also regularly hosts live-streams with guest speakers like local politicians, congressional candidates, and community organizers, all of whom instruct users on how to best agitate for change armed with the data visualizations shared in the group. \textit{``You’re a mom up the street, but you’re not powerless,''} emphasized one of the guest speakers. \textit{``Numbers matter! What is just and what is true matters.} [...] \textit{Go up and down the ladder---start real local. Start with the lesser magistrates, who are more accessible, easier to reach, who will make time for you.''} (July 23, 2020) 

These groups have been incredibly effective at galvanizing a network of engaged citizens towards concrete political action. Local officials have relied on data narratives generated in these groups to call for a lawsuit against the Ohio Department of Health (July 20, 2020). In Texas, a coalition of mayors, school board members, and city council people investigated the state’s COVID-19 statistics and discovered that a backlog of unaudited tests was distorting the data, prompting Texas officials to employ a forensic data team to investigate the surge in positive test rates~\cite{carroll_texas_2020}. \textit{``There were over a million pending assignments} [that were distorting the state’s infection rate],'' the city councilperson said to the group’s 40,000+ followers. \textit{``We just want to make sure that the information that is getting out there is giving us the full picture.''} (August 17, 2020) Another Facebook group solicited suggestions from its followers on how to support other political groups who need data to support lawsuits against governors and state health departments. \textit{``If you were suddenly given access to all the government records and could interrogate any official,''} a group administrator asked, \textit{``what piece of data or documentation would you like to inspect?''} (September 11, 2020) The message that runs through these threads is unequivocal: that data is the only way to set fear-bound politicians straight, and using better data is a surefire way towards creating a safer community. 


\section{Discussion} 

Anti-maskers have deftly used social media to constitute a cultural and discursive arena devoted to addressing the pandemic and its fallout through practices of data literacy. Data literacy is a quintessential criterion for membership within the community they have created. The prestige of both individual anti-maskers and the larger Facebook groups to which they belong is tied to displays of skill in accessing, interpreting, critiquing, and visualizing data, as well as the pro-social willingness to share those skills with other interested parties. This is a community of practice \cite{wenger_communities_1998,lave_situated_1991} focused on acquiring and transmitting expertise, and on translating that expertise into concrete political action. Moreover, this is a subculture shaped by mistrust of established authorities and orthodox scientific viewpoints. Its members value individual initiative and ingenuity, trusting scientific analysis only insofar as they can replicate it themselves by accessing and manipulating the data firsthand. They are highly reflexive about the inherently biased nature of any analysis, and resent what they view as the arrogant self-righteousness of scientific elites.

As a subculture, anti-masking amplifies anti-establishment currents pervasive in U.S. political culture. Data literacy, for anti-maskers, exemplifies distinctly American ideals of intellectual self-reliance, which historically takes the form of rejecting experts and other elites~\cite{hofstadter_anti-intellectualism_1966}. The counter-visualizations that they produce and circulate not only challenge scientific consensus, but they also assert the value of independence in a society that they believe promotes an overall de-skilling and dumbing-down of the population for the sake of more effective social control~\cite{tripodi_searching_nodate,hochschild_strangers_2016,elisha_moral_2011}. As they see it, to counter-visualize is to engage in an act of resistance against the stifling influence of central government, big business, and liberal academia. Moreover, their simultaneous appropriation of scientific rhetoric and rejection of scientific authority also reflects longstanding strategies of Christian fundamentalists seeking to challenge the secularist threat of evolutionary biology~\cite{bielo_particles--peoplemolecules--man_2019}.

So how do these groups diverge from scientific orthodoxy if they are using the same data? We have identified a few sleights of hand that contribute to the broader epistemological crisis we identify between these groups and the majority of scientific researchers. For instance, they argue that there is an outsized emphasis on \emph{deaths} versus cases: if the current datasets are fundamentally subjective and prone to manipulation (e.g., increased levels of faulty testing, asymptomatic vs. symptomatic cases), then deaths are the only reliable markers of the pandemic’s severity. Even then, these groups believe that deaths are an additionally problematic category because doctors are using a COVID diagnosis as the main cause of death (i.e., people who die because of COVID) when in reality there are other factors at play (i.e., dying with but not because of COVID). Since these categories are fundamentally subject to human interpretation, especially by those who have a vested interest in reporting as many COVID deaths as possible, these numbers are vastly over-reported, unreliable, and no more significant than the flu. 

Another point of contention is that of lived experience: in many of these cases, users do not themselves know a person who has experienced COVID, and the statistics they see on the news show the severity of the pandemic in vastly different parts of the country. Since they do not see their experience reflected in the narratives they consume, they look for hyperlocal data to help guide their decision-making. But since many of these datasets do not always exist on such a granular level, this information gap feeds into a larger social narrative about the government’s suppression of critical data and the media’s unwillingness to substantively engage with the subjectivity of coronavirus data reporting.

Most fundamentally, the groups we studied believe that \textbf{science is a process, and not an institution}. As we have outlined in the case study, these groups mistrust the scientific establishment (``Science'') because they believe that the institution has been corrupted by profit motives and politics. The knowledge that the CDC and academics have created cannot be trusted because they need to be subject to increased doubt, and not accepted as consensus. In the same way that climate change skeptics have appealed to Karl Popper’s theory of falsification to show why climate science needs to be subjected to continuous scrutiny in order to be valid \cite{fischer_knowledge_2019}, we have found that anti-mask groups point to Thomas Kuhn’s \textit{The Structure of Scientific Revolutions} to show how their anomalous evidence---once dismissed by the scientific establishment---will pave the way to a new paradigm (\textit{``As I've recently described, I'm no stranger to presenting data that are inconsistent with the narrative. It can get ugly. People do not give up their paradigms easily.} [...] \textit{Thomas Kuhn wrote about this phenomenon, which occurs repeatedly throughout history. Now is the time to hunker down. Stand with the data,''} August 5, 2020). For anti-maskers, valid science must be a process they can critically engage for themselves in an unmediated way. \emph{Increased} doubt, not consensus, is the marker of scientific certitude. 

Arguing that anti-maskers simply need more scientific literacy is to characterize their approach as uninformed and inexplicably extreme. This study shows the opposite: users in these communities are deeply invested in forms of critique and knowledge production that they recognize as markers of scientific expertise. If anything, anti-mask science has extended the traditional tools of data analysis by taking up the theoretical mantle of recent critical studies of visualization~\cite{dignazio_data_2020,correll_ethical_2019}. Anti-mask approaches acknowledge the subjectivity of how datasets are constructed, attempt to reconcile the data with lived experience, and these groups seek to make the process of understanding data as transparent as possible in order to challenge the powers that be. For example, one of the most popular visualizations within the Facebook groups we studied were unit visualizations, which are popular among anti-maskers and computer scientists for the same reasons: they provide more information, better match a reader’s mental model, and they allow users to interact with them in new and more interesting ways \cite{park_atom_2018}. Barring tables, they are the most unmediated way to interact with data: one dot represents one person. 

Similarly, these groups' impulse to mitigate bias and increase transparency (often by dropping the use of data they see as ``biased'') echoes the organizing ethos of computer science research that seeks to develop ``technological solutions regarding potential bias'' or ``ground research on fairness, accountability, and transparency''~\cite{association_for_computing_machinery_acm_2020}. In other words, these groups see themselves as engaging deeply within multiple aspects of the scientific process---interrogating the datasets, analysis, and conclusions---and still university researchers might dismiss them in leading journals as  ``scientifically illiterate''~\cite{miller_science_2020}. In an interview with the Department of Health and Human Services podcast, even Anthony Fauci (Chief Medical Advisor to the US President) noted: \textit{``one of the problems we face in the United States is that unfortunately, there is a combination of an anti-science bias} [...] \textit{people are, for reasons that sometimes are, you know, inconceivable and not understandable, they just don't believe science''}~\cite{fauci_dr_2020}.

We use Dr. Fauci's provocation to illustrate how understanding the way that anti-mask groups think about science is crucial to grappling with the contested state of expertise in American democracy. In a study of Tea Party supporters in Louisiana, Arlie Russell Hochschild \cite{hochschild_strangers_2016} explains the intractable partisan rift in American politics by emphasizing the importance of a ``deep story'': a subjective prism that people use in order to make sense of the world and guide the way they vote. For Tea Party activists, this deep story revolved around anger towards a federal system ruled by liberal elites who pander to the interests of ethnic and religious minorities, while curtailing the advantages that White, Christian traditionalists view as their American birthright. We argue that the anti-maskers' deep story draws from similar wells of resentment, but adds a particular emphasis on the usurpation of scientific knowledge by a paternalistic, condescending elite that expects intellectual subservience rather than critical thinking from the lay public. 

To be clear, we are not promoting these views. Instead, we seek to better understand how data literacy, as a both a set of skills and a moral virtue championed within academic computer science, can take on distinct valences in different cultural contexts. A more nuanced view of data literacy, one that recognizes multiplicity rather than uniformity, offers a more robust account of how data visualization circulates in the world. This culturally and socially situated analysis demonstrates why increasing access to raw data or improving the informational quality of data visualizations is not sufficient to bolster public consensus about scientific findings. Projects that examine the cognitive basis of visualization or seek to make ``better'' or ``more intuitive'' visualizations \cite{kosara_empire_2016} will not meaningfully change this phenomenon: anti-mask protestors already use visualizations, and do so extremely effectively. Moreover, in emphasizing the politicization of pandemic data, our account helps to explain the striking correlation between practices of counter-visualization and the politics of anti-masking. For members of this social movement, counter-visualization and anti-masking are complementary aspects of resisting the tyranny of institutions that threaten to usurp individual liberties to think freely and act accordingly. 


\section{Implications and conclusion} 

This paper has investigated anti-mask counter-visualizations on social media in two ways: quantitatively, we identify the main types of visualizations that are present within different networks (e.g., pro- and anti-mask users), and we show that anti-mask users are prolific and skilled purveyors of data visualizations. These visualizations are popular, use orthodox visualization methods, and are promulgated as a way to convince others that public health measures are unnecessary. In our qualitative analysis, we use an ethnographic approach to illustrate how COVID counter-visualizations actually reflect a deeper epistemological rift about the role of data in public life, and that the practice of making counter-visualizations reflects a participatory, heterodox approach to information sharing. Convincing anti-maskers to support public health measures in the age of COVID-19 will require more than ``better'' visualizations, data literacy campaigns, or increased public access to data. Rather, it requires a sustained engagement with the social world of visualizations and the people who make or interpret them. 

While academic science is traditionally a system for producing knowledge within a laboratory, validating it through peer review, and sharing results within subsidiary communities, anti-maskers reject this hierarchical social model. They espouse a vision of science that is radically egalitarian and individualist. This study forces us to see that coronavirus skeptics champion science as a personal practice that prizes rationality and autonomy; for them, it is \emph{not} a body of knowledge certified by an institution of experts. Calls for data or scientific literacy therefore risk recapitulating narratives that anti-mask views are the product of individual ignorance rather than coordinated information campaigns that rely heavily on networked participation. Recognizing the \emph{systemic} dynamics that contribute to this epistemological rift is the first step towards grappling with this phenomenon, and the findings presented in this paper corroborate similar studies about the impact of fake news on American evangelical voters~\cite{tripodi_searching_nodate} and about the limitations of fact-checking climate change denialism~\cite{fischer_knowledge_2019}. 

Calls for media literacy---especially as an ethics smokescreen to avoid talking about larger structural problems like white supremacy---are problematic when these approaches are deficit-focused and trained primarily on individual responsibility. Powerful research and media organizations paid for by the tobacco or fossil fuel industries~\cite{proctor_golden_2011,oreskes_merchants_2010} have historically capitalized on the skeptical impulse that the ``science simply isn't settled,'' prompting people to simply ``think for themselves'' to horrifying ends. The attempted coup on January 6, 2021 has similarly illustrated that well-calibrated, well-funded systems of coordinated disinformation can be particularly dangerous when they are \emph{designed} to appeal to skeptical people. While individual insurrectionists are no doubt to blame for their own acts of violence, the coup relied on a collective effort fanned by people questioning, interacting, and sharing these ideas with other people. These skeptical narratives are powerful because they resonate with these these people's lived experience and---crucially---because they are posted by influential accounts across influential platforms. 

Broadly, the findings presented in this paper also challenge conventional assumptions in human-computer interaction research about who imagined users might be: visualization experts traditionally design systems for scientists, business analysts, or journalists. Researchers create systems intended to democratize processes of data analysis and inform a broader public about how to use data, often in the clean, sand-boxed environment of an academic lab. However, this literature often focuses narrowly on promoting expressivity (either of current or new visualization techniques), assuming that improving visualization tools will lead to improving public understanding of data. This paper presents a community of users that researchers might not consider in the systems building process (i.e., supposedly ``data illiterate'' anti-maskers), and we show how the binary opposition of literacy/illiteracy is insufficient for describing how orthodox visualizations can be used to promote unorthodox science. Understanding how these groups skillfully manipulate data to undermine mainstream science requires us to adjust the theoretical assumptions in HCI research about how data can be leveraged in public discourse. 

What, then, are visualization researchers and social scientists to do? One step might be to grapple with the social and political dimensions of visualizations at the \emph{beginning}, rather than the end, of projects~\cite{correll_ethical_2019}. This involves in part a shift from positivist to interpretivist frameworks in visualization research, where we recognize that knowledge we produce in visualization systems is fundamentally ``multiple, subjective, and socially constructed''~\cite{meyer_criteria_2019}. A secondary issue is one of uncertainty: Jessica Hullman and Zeynep Tufekci (among others) have both showed how \emph{not} communicating the uncertainty inherent in scientific writing has contributed to the erosion of public trust in science~\cite{hullman_pursuit_2019,tufekci_opinion_2020-1}. As Tufekci demonstrates (and our data corroborates), the CDC's initial public messaging that masks were ineffective---followed by a quick public reversal---seriously hindered the organization's ability to effectively communicate as the pandemic progressed. As we have seen, people are not simply passive consumers of media: anti-mask users \emph{in particular} were predisposed to digging through the scientific literature and highlighting the uncertainty in academic publications that media organizations elide. When these uncertainties did not surface within public-facing versions of these studies, people began to assume that there was a broader cover-up~\cite{tufekci_opinion_2020}.

But as Hullman shows, there are at least two major reasons why uncertainty \emph{hasn't} traditionally been communicated to the public~\cite{hullman_why_2020}. Researchers often do not believe that people will understand and be able to interpret results that communicate uncertainty (which, as we have shown, is a problematic assumption at best). However, visualization researchers also do not have a robust body of understanding about \emph{how}, and when, to communicate uncertainty (let alone how to do so effectively). There are exciting threads of visualization research that investigate how users' interpretive frameworks can change the overarching narratives they glean from the data~\cite{hullman_visualization_2011,peck_data_2019,segel_narrative_2010}. Instead of championing absolute certitude or objectivity, this research pushes us to ask how scientists and visualization researchers alike might express uncertainty in the data so as to recognize its socially and historically situated nature. 

In other words, our paper introduces new ways of thinking about ``democratizing'' data analysis and visualization. Instead of treating increased adoption of data-driven storytelling as an unqualified good, we show that data visualizations are not simply tools that people use to understand the epidemiological events around them. They are a battleground that highlight the contested role of expertise in modern American life. 


\begin{acks}
The authors thank Stephan Risi, Maeva Fincker, and Mateo Monterde for their assistance with quantitative methods and supplemental material. We also thank the members of the Visualization Group (especially Jonathan Zong, Alan Lundgard, Harini Suresh, EJ Sefah, and the Fall 2020 UROP cohort: Anna Arpaci-Dusseau, Anna Meurer, Ethan Nevidomsky, Kat Huang, and Soomin Chun). This manuscript benefited from the insights of Rodrigo Ochigame, Hannah LeBlanc, Meghan Kelly, Will Deringer, Blakeley H. Payne, Mariel García-Montes, and the comments of anonymous reviewers. This project was supported by NSF Award 1900991, NSF Dissertation Improvement Grant 1941577, an SSRC Social Data Dissertation Fellowship, and the MIT Programs for Digital Humanities. 
\end{acks}


\bibliographystyle{ACM-Reference-Format}
\bibliography{covid-vis.bib}

\end{document}